\documentclass[letterpaper,10pt]{article}
\usepackage{osameet2}
\usepackage{amsmath,amssymb}
\usepackage[pdftex,colorlinks=true,bookmarks=false,citecolor=blue,urlcolor=blue]{hyperref} 
\usepackage{soul}
\usepackage{subfigure}

\begin{document}

\title{Evolution towards Smart Optical Networking: Where Artificial Intelligence (AI) meets the World of Photonics}

\author{Admela Jukan and Mohit Chamania}
\address{Technische Universit\"at Carolo-Wilhelmina zu
Braunschweig and ADVA Optical Networking}
\email{a.jukan@tu-bs.de, mchamania@advaoptical.com}

\begin{abstract}
Smart optical networks are the next evolution of programmable networking and programmable automation of optical networks, with human-in-the-loop network control and management. The paper discusses this evolution and the role of Artificial Intelligence (AI).\end{abstract}


\section{Introduction}

\par Smart systems encompass a wide area of applications in computing and sensing that have permeated multiple industrial sectors, such as (smart) cities, transportation, energy and homes. While the term \emph{smart} alone may have a broad meaning, the key differentiator for modern compute and sensing systems over any previous generation is their ability to connect over a \emph{network} with cloud and storage systems. In this way, large data collection, processing, exchange and analysis have become an integral part of a smart system, unlike stand-alone systems of the past. This has naturally led to the emergence of artificial intelligence (AI) as an integral part. AI enables a computing system to implement cognitive functions akin to humans, and its innate abilities to provide solutions to problems such as self-configuration, self-healing, and self-optimization, have propelled it to one of the most promising concepts in networking today.  

\par Two evolving capabilities, namely programmability and elasticity, can make the integrating of AI with optical networks transformative.  \emph{Programmablity} of optical network technologies today leverage the basic concepts of software-defined networking, which decouple the network control plane from the underlying hardware, or data plane. While many systems in the past pioneered the SDN conceptually in the research communities \cite{oscars}, it is not until SDN gained wide industrial acceptance, that the optical layer embraced true software programmability, including open source development projects, such as the Open Network Operating System (ONOS) \cite{onos}, OpenDaylight platform (ODL) \cite{odl}, Open Roadm\cite{openroadm}, and OpenConfig  \cite{openconfig}. \emph{Elasticity} of optical networks is another important technology evolution towards flexible spectrum management. Elastic or flexi-grid optical networks divide optical spectrum into flexible grids (slices), offering flexible and just-enough spectrum to variable bandwidth demands which previous generation of optical networks based on Wavelength Division Multiplexing (WDM) could not offer. This trend has led to the emergence of Sliceable Bandwidth Variable Transponder (SBVT) technologies that on its own provides programmability deeper in the photonics, making it possible to choose among various modulation schemes, spectrum slice, including tunability of physical layer parameters. It is intuitive to envision that adding the AI capabilities to the concepts of elasticity and programmability is expected to result in a transformative ability of optical network to serve future applications.

 \par The paper discusses the evolution of optical networking towards \emph{smart optical networks} enabling a seamless integration of Artificial Intelligence (AI). We first revisit the concepts of programmable networking and postulate that its evolution towards programmable automation of optical networks is the first step towards AI. We then present the vision of smart optical networks with their integral concepts of programmability, elasticity and AI. 

\section{From Programmable Networking to Programmable Automation of Networks}
\par Optical network systems have been traditionally designed to enable Internet traffic to grow smoothly, without major aspirations to support fully automated IP-optical or other cross-layer programmability. 
Figure 1a depicts a state-of-the-art management system for optical networks. Network control orchestration is the primary focus of programmability in current networks, with a special focus on automation of network control operations such as provisioning of services. Approaches that focus on the control of individual devices, typical to IP networks, have not gained traction in optical networks, due to the challenges with modeling the physical properties of light propagation while working with abstracted device representations. As a result, programmable network control has focused on definition of operations at the network scope \cite{tapi}, while leaving the control of individual domains to vendor-specific softwares. 

\par At the same time, automation of network management operations is a relatively unexplored field. The primary challenge in the context of optical network management is related to the complexity associated with different physical components involved in optical networking (AWGs, amplifiers, WSSs, lasers) in different configurations and the vendor specific representations for these components. There has been work to employ AI to interpret network management data to perform root-cause analysis of failures and even predict failures in the network, but has typically been siloed within vendor implementations and is not accurate enough to date. As a result, these tools serve as a potential diagnostic tool, but network management operations still require interpretation and intervention from experienced network operators. The operation of a network also includes significant interaction with other software subsystems which is has a strong institutional dependency, and is typically a part of the Operations Support System (OSS). 


\par While there have been significant advances in programmable networking, full automation of network operations has not achieved widespread adoption. The primary challenge has been the lack of mechanisms to incorporate humans-in-the-loop, which is an intrinsic requirement.  Also, as seen in Fig. 1a, reliance on humans to perform diagnostics and analysis during management operations, and take appropriate actions has limited the scope of automation to network control. Finally, most techniques for automation have not addressed the challenge of integrating these systems with numerous external subsystems (billing, reporting etc.) which is typically under the purview of the OSS.

\par Figure 1b illustrates the evolution which can address the challenges of programmable networking, as we envision it. \emph{Orchestration} is a key component in this architecture, and has been used to reduce the number of human operations, and consequently the operational complexity, of typical operations. Numerous methods have been proposed to create programmable workflows that reflect basic services in infrastructure management: Programmable workflows allow users to define \emph{how} to orchestrate network operations, with the syntax of the workflow defined using traditional programming languages \cite{oscars} or as abstract workflows \cite{bpel}. The so-called network operating systems \cite{onos,odl} have also experimented with the use of intent-based interfaces, providing users a syntax to define \emph{what} is desired from the network, and translating these commands into workflows to orchestrate network operations. As seen in Figure 1b, orchestration is envisioned as a key enabler for incorporating programmable automation, while facilitating human intervention if desired. Cognitive management systems will observe and learn from human operators to understand the fundamental network-centric as well as other business-related operations and eventually automate these operations while still potentially allowing for human intervention and correction. The network operators, as Figure 1b illustrates, are still an integral part of the future systems, and are now able to chose the level of automation, or override the compilers of intent-based instruction, which otherwise would run fully automatically. AI-based cognitive network and management is envisioned, which can to talk to humans in human (or at least intent-based) languages, while addressing the problems at scale that only machines can. At the same time, automation of network control will be enhanced by AI-based techniques to perform optimization and traditional network management operations requiring operator intervention, primarily for diagnostics, will be replaced by cognitive techniques. 

  \begin{figure}[t]
    \centering
      \subfigure[Optical networks today] {
        \includegraphics[width=0.47\columnwidth, trim={0 3.8cm 12cm 0}, clip]{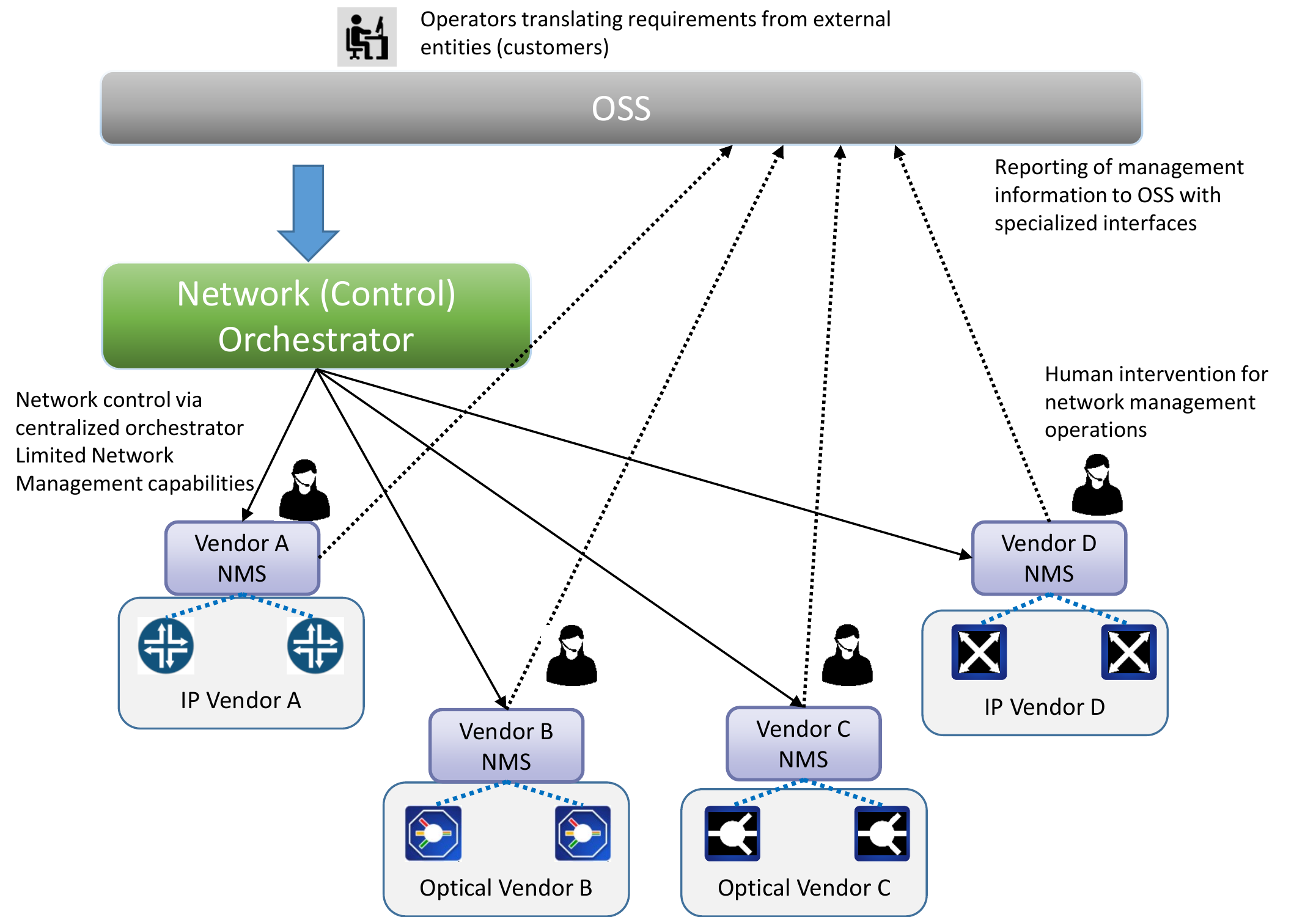}
        \label{fig1a}
      }
      \subfigure[Future optical networks] {
        \includegraphics[width=0.47\columnwidth, trim={0 3.8cm 12cm 0}, clip]{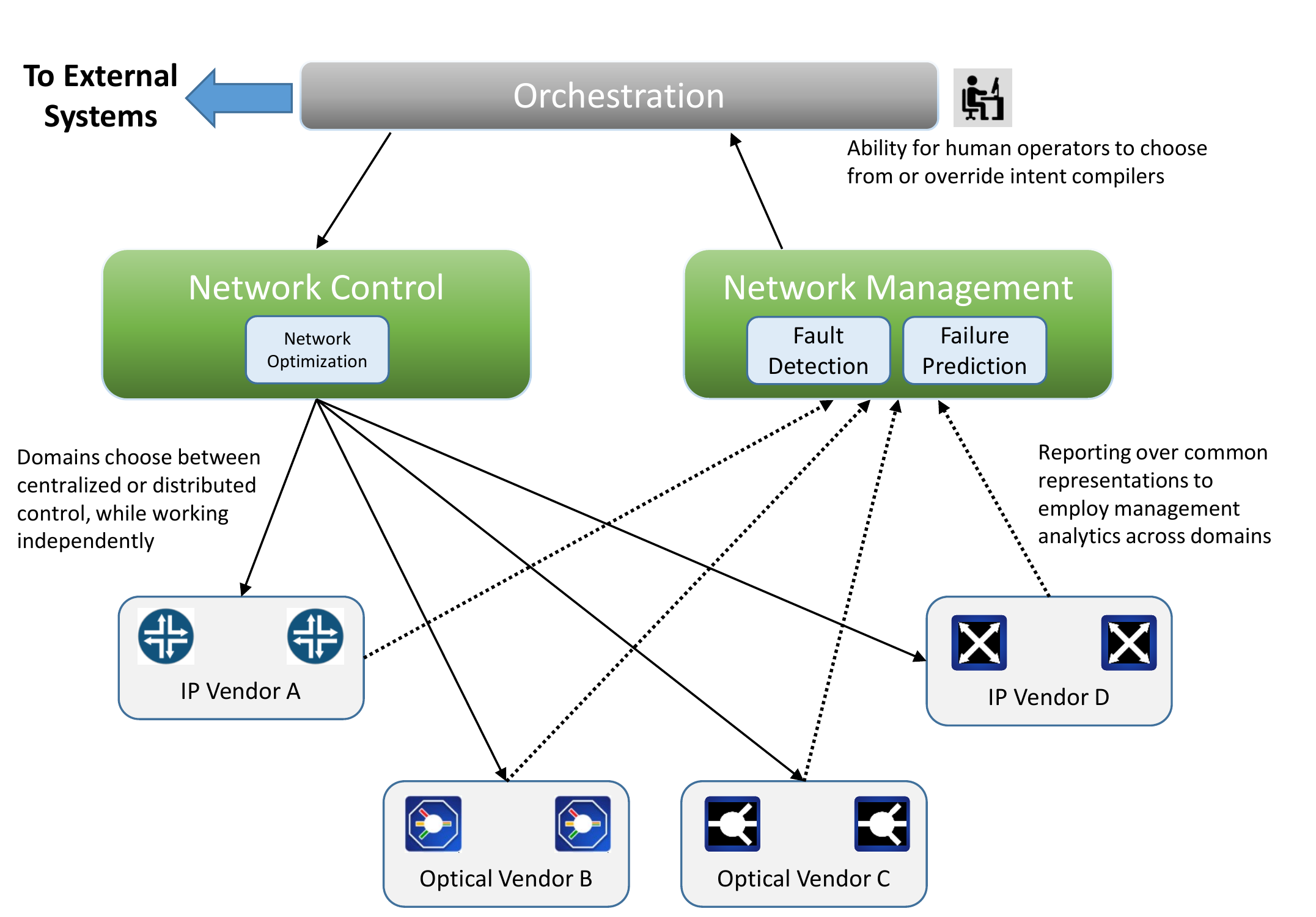}
        \label{fig1b}
      }
    \vspace {-0.3cm}
    \caption{\small Evolution of optical networking towards AI-based cognitive network and management.}
    \vspace {-0.8cm}
\end{figure}

\section{The Role of AI: Challenges and Opportunities}

\par To evolve towards the vision of AI-enabled optical networks shown in Figure 2b, we see three main areas of influence. 

\subsection{Network Control and Optimization} 
\par Network control has been extensively researched, with areas focusing primarily on interface specifications for programmatic control and optimization for basic network control operations. With the introduction of automation in the network, and growing demand for capacity, network control will evolve to incorporate complex network optimization operations. Genetic Algorithms (GA) are one of the popular algorithms in optical networking among many evolutionary algorithms discovered and designed for optimizing complex systems, which has been used extensively and in a myriad network design problems. Swarm based systems have also been used to solve network optimization problems: \cite{Kyriakopoulos2016} proposed a heuristic method based on ant colony optimization to reduce network energy footprint, whereas \cite{Largo2012}  presented a comparative study among three multi-objective evolutionary algorithms (MOEAs) based on swarm intelligence to solve the RWA problem in optical networks. We expect that the increasing bandwidth requirements will demand network optimization as a core network feature, while the introduction of flex-grid optics, sliceable variable bandwidth transponders and other emerging optical technologies will further increase the computational complexity of the optimization problem. AI-based techniques will play a major role for optimization mechanisms to support the scale and flexibility demanded from the next generation of programmable photonic infrastructure.

\subsection{Network Management}
\par Research in the use of AI for Network Management operations has focused on the identification/classification of problems related to the optical channels.  Machine Learning algorithms have been used to address the issue of nonlinearity in optical fiber channel, and also in extracting information about the optical signal detection \cite{Zibar2017}. Neural networks can be used OSNR estimation and modulation format classification \cite{Thrane2017}. A loss classification technique for OBS networks based on machine learning techniques was proposed in \cite{Jayaraj2008}. This approach used both a supervised learning technique (hidden Markov model (HMM)) and an unsupervised learning technique (expectation maximization (EM) clustering) on the observed losses and classified them into a set of states (clusters) after which an algorithm differentiates between the congestion and contention losses. Machine learning algorithms based on an artificial neural network were also used to provide robust and adaptive traffic models and cognitive receiver design \cite{Morales2017, Borkowski2015}. 

\par One of the big challenges with network management is the collection and interpretation of data from vendor specific devices. Data collection in optical networks is primarily based on SNMP which has vendor specific representations and is inefficient for collecting measurement data. To this end, projects like OpenROADM which attempt to define management models for the various entities used in an optical network and concepts like streaming telemetry, introduced in OpenConfig, are key enablers for efficient data collection, which in turn will make AI-based algorithms more portable in multi-vendor environments. In conjunction, AI based techniques will become more prevalent in replacing human operators for interpreting data, and applications like fault management, failure prediction and intrusion detection will be likely driven by cognitive systems. It is also clear that the models of choice for network control and management today also have limited overlap: models that abstract complexity of device configuration are favored for network control operations, while a more detailed model is desired for management and monitoring operations. Automation of network operations will also require interpreting cues from the network (e.g. failure prediction) which might potentially lead to re-configuration of the network, and consequently, mechanisms to maintain and correlate multiple \emph{views} of the network based on different models will be essential for automation of network operations.  

\subsection{Intent-based API and Orchestration}
\par Orchestration of operations requires significant operator involvement in current networks, and we see this as the prime target to be enhanced using techniques derived from AI. The future orchestration subsystem is illustrated in Figure 2. Requests to the orchestration engine will be defined via an intent-based interface, and integration over other interfaces (including voice based, text based or even s/w based) would involve subsystems that can interpret the \emph{intent} of an incoming request, much like the techniques used in Natural Language Processing to date.  

\par After receiving an intent, the orchestration will try and serve an intent, which could be performed via pre-programmed intent compilers. In case a suitable pre-programmed compiler is not available, we envision the use of AI to generate workflows to serve an intent on the fly. Research in computer science has explored the use of AI techniques to generate \emph{code} for a defined problem \cite{deepcoder}. With advances in this area, we envision that such algorithms would be able to break-down complex requirements defined in an intent and create workflows for the same. At the same time, based on the experience with AI-based techniques to date, there will likely be no single mechanism to generate the best workflows for all situations, and we expect that a number of such techniques will work in parallel and give multiple potential solutions. 
Intent negotiation, seen in Figure 2a, provides the mechanism to choose from one of the potential solutions generated. The negotiation framework receives an incoming intent request, and broadcasts it to all suitable components, and has the capability to choose from a set of potential workflows for a given intent. The intent negotiation block also provides a uniquely suitable point for human intervention, allowing an operator to run the orchestration to run in a fully automated fashion, review and choose from potential workflows proposed, or even override with a human-defined workflow. Figure 2b illustrate a human operator talking to a AI-enabled assistant about scheduling a secure city cloud service for a major city marathon. The network operator is part of the AI-enabled system embedded into the network and cloud infrastructure of the future. The network operator is "gone fishing", while asking the network operator (avatar) to "send configuration and logs to her mobile phone".
  
\section{Conclusion}
\par  We expect the automation of optical network operations to be replaced by full AI-based cognitive network and management, which can to talk to humans in human (or at least intent-based) languages, while addressing the problems at scale that only machines can.The human-in-the-loop systems will retain human network operators as an integral part of photonic systems, able to chose the level of automation, or override the compilers of intent-based instruction, which otherwise would run fully automatically. We also expect optimization and management operations to empoy AI-based techniques to adapt to the demands of the next generation of programmable photonic infrastructure.

  \begin{figure}[t]
                   \vspace {-2cm}
  \centering
      \subfigure[AI-enabled orchestration and intent] {
        \includegraphics[width=0.36\columnwidth, trim={0 1.5cm 0 7.2cm}, clip]{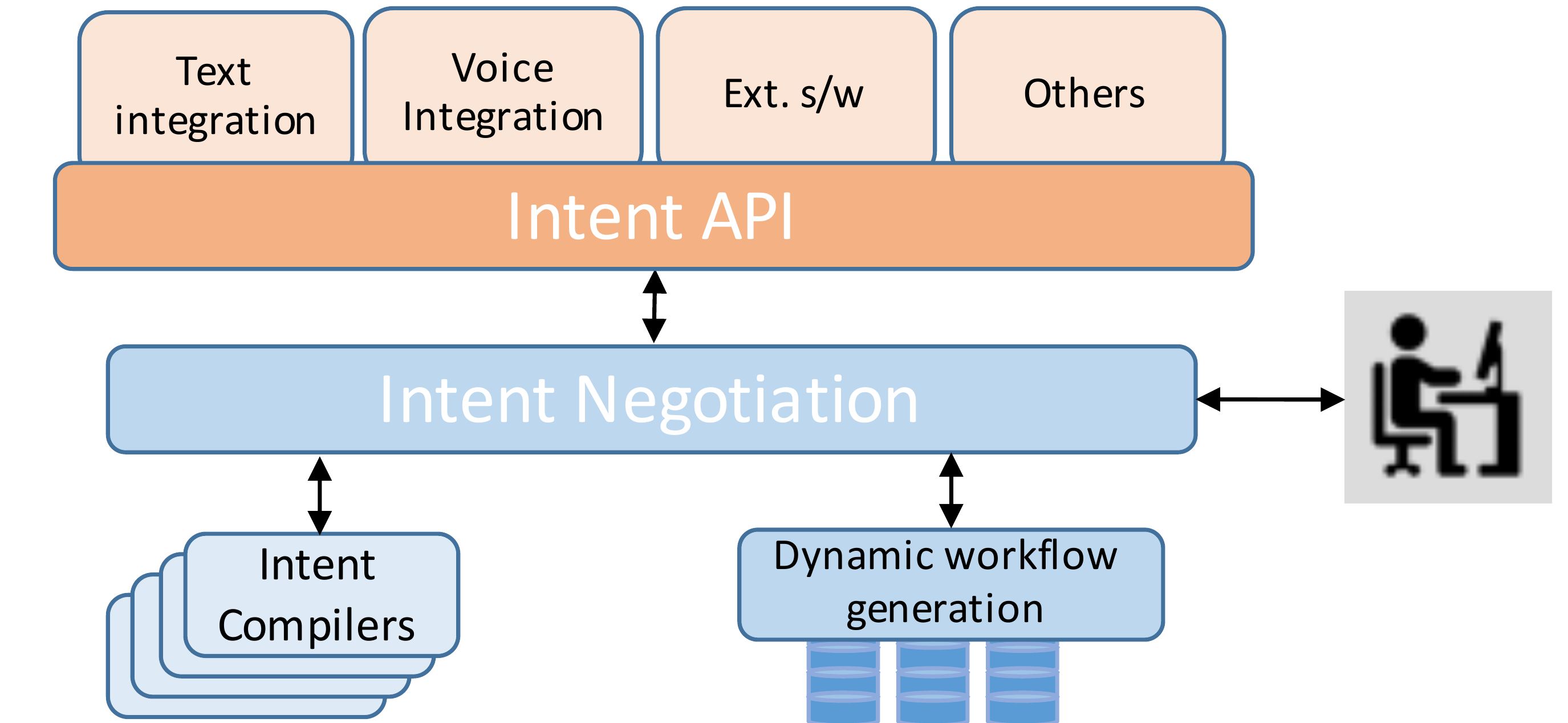}
           \vspace {-2cm}
\label{fig2a}
      }
      \subfigure["Smart City Operator 2030"] {

        \includegraphics[width=0.35\columnwidth, trim={4cm 1cm 2.5cm 6cm}, clip]{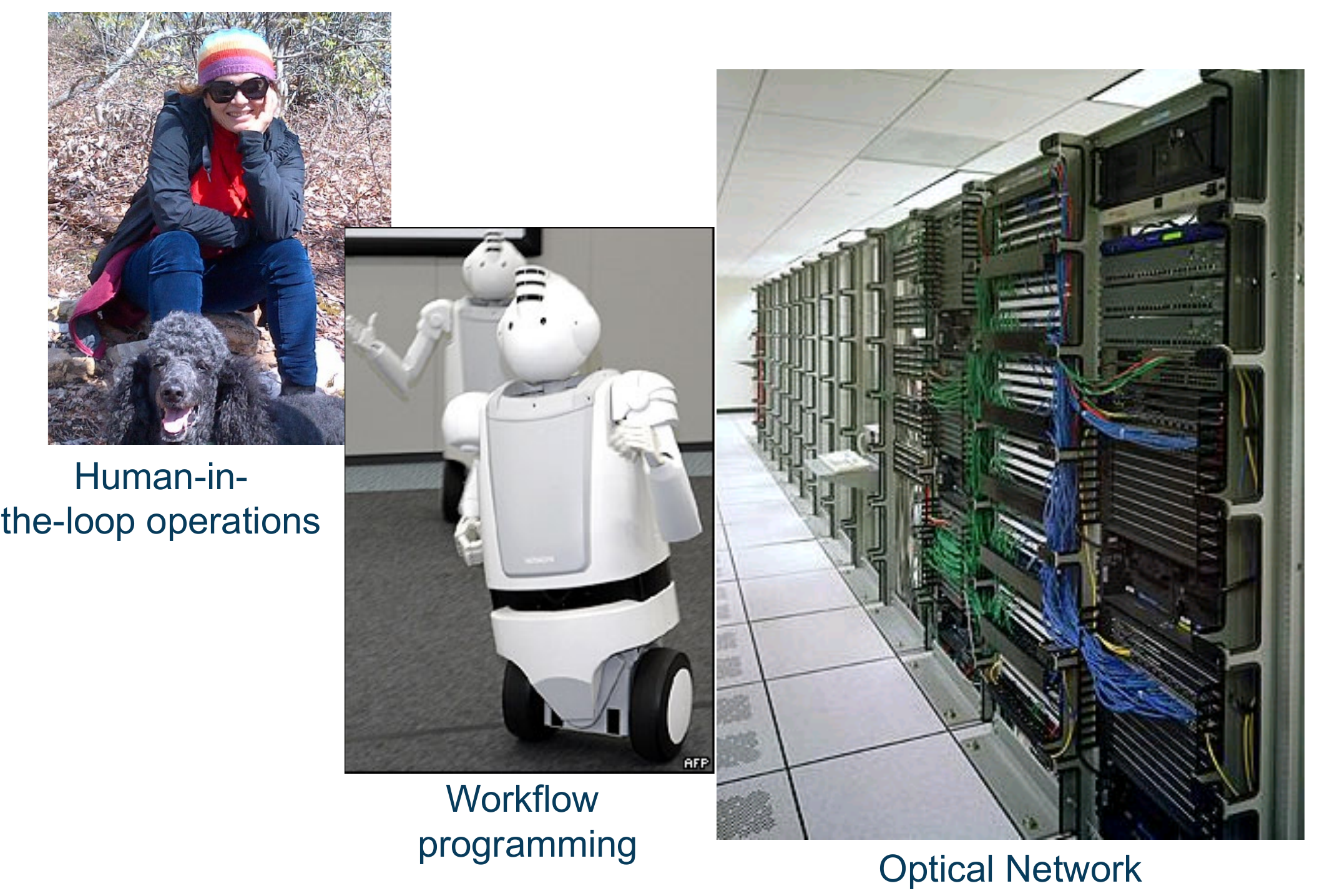}
     \label{fig2b}
      }
    \vspace {-0.3cm}
    \caption{\small Illustrative examples of intent-based AI and automatic orchestration}
     \vspace {-0.8cm}
\end{figure}

\end{document}